\documentclass{aa}
\usepackage{graphics}
\begin{document}

   \thesaurus{		 
              (10.19.3;  
               13.18.1;  
               09.08.1;  
               09.19.2)}  
   \title{The radio emission from the Galaxy at 22 MHz}

   \author{R. S. Roger
          \inst{1}
           \and
           C. H. Costain 
	  \inst{1,2}
	   \and
	   T. L. Landecker
          \inst{1}
	   \and
	   C. M. Swerdlyk
	\inst{1,3}
          }

   \offprints{R. S. Roger}

   \institute{National Research Council Canada, Herzberg Institute of
Astrophysics, Dominion Radio Astrophysical Observatory, Penticton, B.C.
Canada  V2A 6K3\\ email: robert.roger@hia.nrc.ca
         \and
   deceased December, 1989
	 \and
   now at University of Victoria, Victoria, B.C., Canada
             }

   \date{Received October 22, 1998; accepted January 7, 1999}

   \maketitle

   \begin{abstract}

We present maps of the 22 MHz radio emission between declinations
$-$28\degr~ and $+$80\degr, covering $\sim$73$\%$ of the sky, derived
from observations with the 22 MHz radiotelescope at the Dominion Radio
Astrophysical Observatory (DRAO).  The resolution of the telescope (EW
$\times$ NS) is 1.1\degr\ $\times$ 1.7\degr\ secant(zenith angle).
The maps show the large scale features of the emission from the Galaxy
including the thick non-thermal disk, the North Polar Spur (NPS) and
absorption due to discrete \ion{H}{ii} regions and to an extended band
of thermal electrons within 40\degr\ of the Galactic centre.  We give
the flux densities of nine extended supernova remnants shown on the
maps.  A comparison of the maps with the 408-MHz survey of Haslam et
al (\cite{haslam82}) shows a remarkable uniformity of spectral index
(T$\propto \nu^{-\beta}$) of most of the Galactic emission, with
$\beta$ in the range 2.40 to 2.55.  Emission from the outer rim of the
NPS shows a slightly greater spectral index than the distributed
emission on either side of the feature.  The mean local synchrotron
emissivity at 22 MHz deduced from the emission toward nearby extended
opaque \ion{H}{ii} regions is
$\sim$1.5$\times$10$^{-40}$Wm$^{-3}$Hz$^{-1}$sr$^{-1}$, somewhat
greater than previous estimates.

   \end{abstract}

%

\section{Introduction}

The nonthermal radio continuum emission from the Galaxy at frequencies
below 100~MHz is the synchrotron radiation of cosmic ray electrons
with energies of order 1 GeV.  Measurements of the emissivity and the
spectral index of the emission provide direct information about the
electron energy spectrum and the magnetic field strength in the
Galaxy.  At frequencies below about 40 MHz, the opacity of the thermal
component of Galactic emission is often sufficient to absorb the much
brighter background synchrotron emission, providing a means of
estimating local synchrotron emissivities in the directions of \ion{H}{ii}
regions at known distances. 

There have been numerous surveys made of Galactic emission below 500
MHz including: the all--sky map of Haslam et al. (\cite{haslam82}) at
408~MHz; the northern sky maps of Turtle \& Baldwin (\cite{turtle}:
178~MHz, $-$5\degr\ to +90\degr), of Milogradov-Turin \& Smith
(\cite{milogradov}: 38~MHz, $-$25\degr\ to +70\degr) and of Caswell
(\cite{caswell}: 10~MHz, $-$6\degr\ to +74\degr, 0$^h$ to 16$^h$); and
southern sky maps of Landecker \& Wielebinski (\cite{landecker}: 150
MHz and 85 MHz, $-$15\degr\ to +15\degr), Hamilton \& Haynes
(\cite{hamilton}: 153 MHz, +5\degr\ to $-$90\degr), Alvarez et al.
(\cite{alvarez}: 45 MHz, +19\degr\ to $-$90\degr), and Ellis
(\cite{ellis}: 16.5 MHz, 0\degr\ to $-$90\degr).

The 22 MHz Telescope at the DRAO was used in the period 1965 to 1969
to measure the emission from discrete sources and to map the
background radiation from our galaxy.  The telescope has been
described completely by Costain et al. (\cite{costain}); only those
details relevant to the present paper are given here.  Flux densities
of point sources have been published by Roger et al. (\cite{roger69b})
and by Roger et al. (\cite{roger86}).  The low-frequency spectra of
these sources have been discussed by Roger et
al. (\cite{roger73}). Technical problems prevented a satisfactory
calibration of the Galactic emission, but these problems have now been
circumvented and in this paper we present a map of the Galactic
emission at 22~MHz between declinations $-$28\degr\ and $+$80\degr\
covering the complete range of right ascension, $\sim$73\%~ of the
sky.  By comparing the 22~MHz map with the 408~MHz data from the
all-sky survey of Haslam et al. (\cite{haslam82}) we have prepared a map
of the spectral index of the emission over the same area. We also
derive values of the local synchrotron emissivity.


\section{The telescope}

The telescope was an unfilled aperture in the form of
a T with dimensions 96$\lambda$ $\times$ 2.5$\lambda$ east-west (EW)
and 32.5$\lambda$ $\times$ 4$\lambda$ north-south (NS). This array
formed a nearly Gaussian beam of extent 1.1 $\times$ 1.7 degrees (EW
$\times$ NS) at the zenith (declination 48.8\degr). The telescope was
operated as a meridian transit instrument, steerable in declination
between $-$30\degr\ and the North Celestial Pole by the adjustment of
the phases of rows of dipole elements in the north-south direction.
Away from the zenith, foreshortening of the array broadened the beam
in this direction to 1.7{\degr}~secant(Z), where Z is the zenith
angle. The gain of the telescope in directions away from the zenith
was further reduced by the response of the basic array element, a
full-wave dipole ${\lambda}\over{8}$ above a reflecting
screen. Aperture grading by means of attenuators was used to keep
sidelobes to a level of a few percent.

The pencil-beam response was formed by multiplying the signals
from the EW and NS arms. A problem in T-- and cross--format
telescopes arises from the region in which the two arms
intersect. If this region is removed, a broad negative response
is produced around the narrow pencil beam, and the telescope
filters out the lowest spatial frequencies, leading to a poor
representation of the broadest angular components of the
emission. To overcome this problem, the signals from the elements
in the overlap region were split and fed to both the EW and NS
arrays. 

The gain of the antenna (in effect the ratio between the flux density 
of a radio source and the antenna temperature it produces in the main 
beam) was established
using an assumed flux density of 29100 $\pm$ 1500 Jy for Cygnus A. 
Details of the original flux-density scale and the subsequent revision
can be found in Roger et al. (\cite{roger73}) and in Roger et al. 
(\cite{roger86}).


\section{Observations and basic data reduction}

Observations were made in several modes during the working life
of the telescope. In the early part of the observation period the
telescope formed only one beam which could be moved in
declination by operating the NS phasing switches.  Long
scans at fixed declination were made to observe the background
emission as well as point sources, and short scans were made to
measure the flux density of point sources with the beam switched
frequently between different declinations. Only the scans of long
duration were used to assemble the data presented in this paper.
At a later stage, more automated phasing switches were added to 
permit rapid time--shared observations at five adjacent
declinations. A large fraction of the present data was obtained
from long scans with this equipment.

Scans were made at a set of standard declinations, chosen to
sample the emission at half-beamwidth intervals. Since the NS
beamwidth increased with increasing zenith angle, the standard
declinations were not evenly spaced.

The observations were made mostly in the years 1965 to 1969 which
covered a period of fairly low solar activity. Nevertheless,
because of the low frequency of operation, the influence of the
ionosphere on the observations was often large. Observations of
point sources were affected by refraction, scintillation, and
absorption in differing degrees. A correction factor for
ionospheric absorption (primarily a daytime phenomenon) was derived from an 
on--site 22 MHz riometer and was applied to the data.
Refraction amounting to a significant fraction of the (EW) beam
was detected only near times of sunrise in the ionosphere when
large horizontal gradients of electron density were present.  
Scintillation
of ``point sources'' (sources less than 15\arcmin\ diameter) was
a frequent occurrence.   Due to lack of correlation of the
phase and amplitude variations over the extent of the telescope
array during times of severe scintillation, flux densities could
be seriously underestimated.  To overcome this problem, many observations
of each source were made and measurements of flux density were derived
from only those observations judged to be the least affected.  
By contrast, most observations of the extended emission were
largely unaffected by scintillation and, after
correction for absorption, could be reliably averaged together.
The data presented here are the result of averaging at least two 
observations, and in many parts of the sky up to five good observations 
contributed to the average.

Thus, the basic data set is a data array assembled from the averaged
scans at the standard declinations.  Because neighbouring averaged scans
contained data observed at different times with varying 
conditions, the array displayed significant scanning artifacts.
These were greatly reduced by the application of a
Fourier filtering process in the declination dimension.  The data were 
then interpolated onto a grid sampled at 1 minute intervals in 
right ascension and 15\arcmin~ in declination.


\section{Further data processing}

Comparison of the resulting map with maps at higher frequencies
suggested that the brightness temperature calibration of the 22 MHz
map had a zenith--angle--dependent error.  We used the 408 MHz all-sky
survey of Haslam et al. (\cite{haslam82}) for comparisons because
these single--parabloid observations would likely be free of such
defects, but comparisons with other similar data sets would probably
have produced similar conclusions.

For comparison with the 22 MHz map we convolved the 408 MHz data from
its original resolution of 51\arcmin~ to the variable beamwidth of our
data. Calculation of a map of spectral index between the two
frequencies indicated a systematic variation of spectral index with
declination.  Since such an effect is unlikely to be real, we
suspected a calibration problem in the 22 MHz data.  Although the
cause is not fully understood, we believe that the response of the
telescope to extended structure differed from its response to point
sources in directions away from the zenith.  This effect is discussed
in Section 6.2.  We believe, however, that the response of the
telescope at the zenith (declination 48.8\degr) {\it{is}} well
understood. At this declination, we plotted brightness temperature at
22 MHz against brightness temperature at 408 MHz (a T-T plot).  We
used data at all right ascensions except (a) near the Galactic plane
in the Cygnus region (at $\sim$21$^h$, 48\degr) where strong
absorption features are evident in the 22 MHz map (see below) and (b)
regions around a few strong small-diameter sources.  Fig.~1
shows this plot.  The highest temperatures in Fig.~1
correspond to the Galactic plane in the anticentre (at 4$^h$ 40$^m$,
48\degr) where there may be a small amount of absorption, causing the
T-T plot to deviate from a straight line. Using all the data shown in
Fig.~1 we derived a differential spectral index of
2.52$\pm$.02. Restricting the fit to regions with $T_{22}<50$~kK
(corresponding to right ascensions between 5$^h$ 40$^m$ and 18$^h$
10$^m$) the correlation is very tight and the differential spectral
index is 2.57$\pm$.02. These values of spectral index are close to the
value expected in this frequency range from the work of Bridle
(\cite{bridle}) and Sironi (\cite{sironi}) indicating that the
temperature scale at 22 MHz is accurate. Furthermore, the
extrapolation of the line fitted to the data points passes close to
the origin of the T-T plot, giving us confidence that the zero level
of the 22 MHz measurements is also well established at the zenith. T-T
plot analyses at other declinations (again excluding areas of
absorption) indicate that the zero level is acceptably correct
throughout the declination range, but that the temperature scale
varies (the accuracy of the zero level is discussed in section 6.1).

The temperature scale at declinations away from the zenith was
adjusted using the following procedure. At each declination the
average brightness temperature ratio between 22 MHz and 408 MHz
was calculated (using T-T plots) over the range 8 to 16 hours in 
right ascension (to exclude extended absorption regions on the 
Galactic plane). The temperature scale at 22 MHz was then adjusted 
to make this ratio equal to that at the zenith (note that Fig.~1
shows data over a wider range of right ascension).

In order to cover the central regions of the Galaxy, we have included
observations as far south as declination $-$28\degr, where the telescope was
operating at a zenith angle of 76.8\degr. At these large zenith angles, there
is some departure of the antenna gain function from its calculated value
(see Costain et al. \cite{costain}). However, reliable flux densities of
point sources have been obtained as far south as declination $-$17.4\degr\
(Roger et al. \cite{roger86}) and we believe that our calibration procedure
remains reasonably effective to the southern limit of our map.


\section{Results}

\subsection{Map of the northern sky at 22 MHz}

Figs.~2 to 6 show in equatorial coordinates a
contoured gray-scale map of the 22 MHz emission from the sky between
declinations $-$28\degr\ and 80\degr\ in five segments. Figs.~7 and 8
depict the data between Galactic latitutes $-$40\degr\ and
$+$40\degr\ with the same contours and grayscale in Galactic
coordinates, and with positions of extended Galactic sources
indicated.  Fig.~9 is a grayscale representation of the full data
set in Aitoff projection of Galactic coordinates.

The brightness temperature of the 22 MHz emission varies from $\sim$17~kK
towards a broad minimum about 50\degr\ off the Galactic plane at the
longitude of the anticentre to over 250~kK on the plane near the 
Galactic centre.  The brightness temperature near both north and 
south Galactic poles is approximately 27~kK.  The Galactic plane itself
is apparent over the full range of longitude from $+$1\degr\ to $+$244\degr.
At various points along the plane, particularly at lesser longitudes,
depressions are apparent in the emission.  These represent thermal
free-free absorption of bright synchrotron background emission by 
relatively nearby, opaque regions of dense ionized gas.

Two other large--scale features apparent in the maps are Loop I, the 
North Polar Spur (NPS), rising from the plane near longitude 30\degr,
and Loop III, centred near longitude 87\degr.   

We emphasize that the main value of the data lies in the
representation of structure larger than the beam. The
strongest point sources (Cas A, Cyg A, Tau A and Vir A) have been
removed from the map. While other point sources remain in
the maps, these data can not be used to determine their flux
densities. First, the ionospheric effects mentioned above cause
the point sources to be very poorly represented in these maps.
Second, the scaling applied after comparison with the 408 MHz
data will have further affected the flux densities of point
sources at declinations away from the zenith. Reliable point
source flux densities are already available in the published
lists referred to in Section 1.

\subsection {Extended Galactic sources -- supernova remnants}

A number of extended supernova remnants are apparent in the data and
the positions of these are indicated with labels in Figs. 7 and 8.
The flux densities of most of the SNRs have been previously measured
from the original observations and published in various papers.  We
have collected these and listed them in Table 1 together with new flux
densities for two additional remnants not previously reported.  One
other SNR, HB21, is indicated in Fig.~8 but {\it not} listed
in Table 1 because of difficulties in separating its emission from
that of nearby confusing sources.

\begin{table*}
\caption[]{Flux densities of supernova remnants at 22~MHz}
\label{tabl-1}
\begin{center}
\begin{tabular}{llll}
\hline
SNR & Designation & Flux density & Reference \\
~ & ~ & Jy & ~ \\
\hline
$G34.7-0.4$ & W44,3C392 & 881$\pm$108 & Roger et al. (\cite{roger86}) \\
$G74.0-8.5$ & Cygnus Loop & 1378$\pm$400 & this paper \\
$G93.3+6.9$ & DA530,4C(T)55.38.1 & 74$\pm$20 & Roger \& Costain 
(\cite{roger76}) \\
$G119.5+10.2$ & CTA1 & 632$\pm$200 & Pineault et al. (\cite{pineault}) \\
$G120.1+1.4$ & 3C10, Tycho SN1572 & 680$\pm$54 & Roger et al. 
(\cite{roger86}) \\
$G132.7+1.3$ & HB3 & 450$\pm$70 & Landecker et al. (\cite{landecker87}) \\
$G160.9+2.6$ & HB9 & 1130$\pm$340 & this paper \\
$G184.6-5.8$ & Crab Nebula, SN1054 & 3160$\pm$380 & Roger et al. 
(\cite{roger86}) \\
$G189.1+3.0$ & IC443, 3C157 & 615$\pm$75 & Roger et al. (\cite{roger86}) \\
\hline
\end{tabular}
\end{center}
\end{table*}

\subsection {Absorption of the background emission -- \ion{H}{ii} regions}

Depressions in the background emission near the Galactic plane are
identified with a number of extended \ion{H}{ii} regions, which at
frequencies near 20~MHz will largely obscure background emission.  We
list the properties of 21 of these discrete absorption regions in
Table 2, with positions plotted in Figs. 7 and 8.

\begin{table*}
\caption[]{\ion{H}{ii} regions in absorption}
\label{tabl-2}
\begin{center}
\begin{tabular}{llll}
\hline
Galactic Coordinates & Region & Other Designation & Sharpless diameter \\
\hline
$6.3+23.5$ & Sh~27 & $\zeta$-Oph & 480\arcmin \\
$18.7+2.0$ & Sh~54 & II~Ser, NGC6604 & 140\arcmin \\
$49.0-0.5$ & Sh~79(?) & & 40\arcmin \\
$59.4-0.1$ & Sh~86 & I~Vul, NGC6820 & 40\arcmin \\
$85.5-1.0$ & Sh~117 & NGC7000 & 240\arcmin \\
$87.6-3.8$ & Sh~119 & & 160\arcmin \\
$98.5+8.0$ & Sh~129 & I~Cep & 140\arcmin \\
$99.3+3.7$ & Sh~131 & I~Cep, IC1396 & 170\arcmin \\
$102.8-0.7$ & Sh~132 & II~Cep & 90\arcmin \\
$108.9+6.1$ & Sh~150(?) & & 40\arcmin \\
$118.5+6.0$ & NGC7822 & IV~Cep, Sh~171(?) & 180\arcmin \\
$134.8+0.9$ & Sh~190 & IC1805, W4 & 150\arcmin \\
$137.6+1.1$ & Sh~199 & IC1848, W5 & 120\arcmin \\
$148.5-0.2$ & Sh~205 & I~Cam & 120\arcmin \\
$160.1-12.3$ & Sh~220 & NGC1499, II~Per & 320\arcmin \\
$172.0-2.2$ & Sh~229 (230?) & IC405, (I~Aur ?) 
  & 65\arcmin (300\arcmin) \\
$195.1-12.0$ & Sh~264 & $\lambda$-Ori & 390\arcmin \\
$202.9+2.2$ & Sh~273 & NGC2264 & 250\arcmin \\
$206.3-2.1$ & Sh~275 & I~Mon, NGC2244 & 100\arcmin \\
$209.0-19.5$ & Sh~281 & Orion Neb, NGC1976 & 60\arcmin \\
$224.5-1.9$ & Sh~296 (292?) & (IC2177?) & 200\arcmin (21\arcmin) \\
\hline
\end{tabular}
\end{center}
\end{table*}

Figs. 2 to 8 and particularly 9 show the
extended trough of absorption between $l=10$\degr~ and $l=40$\degr.
This trough undoubtedly extends to and past the Galactic centre but
the increasingly extended N-S width of the telescope beam at large
zenith angles was unable to fully resolve the feature below
$l=10$\degr.
 
\subsection {The spectral index of emission, 22 to 408 MHz}

Fig.~10 shows a map of spectral index calculated from the final 22
MHz map and the 408 MHz map (Haslam et al. \cite{haslam82}), the latter
convolved to the declination--dependent beamwidth of the 22 MHz
telescope.  The spectral index, $\beta$, as displayed, is related to
the brightness temperatures at each frequency, $T_{22}$ and $T_{408}$,
by the expression \[\beta = log (T_{22} / T_{408}) / log (408/22). \]
Because the 408 MHz map has been used to establish the variation of
the 22 MHz temperature scale with declination, great care is needed in
interpreting this map.  The process of revising the 22 MHz scale could
eliminate or reduce spectral index features between 8 and 16 hours
right ascension with structure in the declination direction if they
extend over a large range in this dimension.  On the other hand,
features in the spectral index map which have structure in the right
ascension dimension are likely to be largely unaffected by the
correction process.  Similarly, spectral index features with structure
in various directions, including most features which have counterparts
in the individual maps, would be suspect only if distortions appeared
in the declination dimension.  No such artefacts are apparent.
However, some ``banding'' in declination, particularly below
$-$3\degr, can be seen in specific right ascension ranges.  This
effect is easily recognized as spurious and is probably related to
zero level errors in the 22~MHz or, possibly, in the 408~MHz map. Such
effects may be aggravated by the large zenith angles at which these
low-declination regions were observed at 22 MHz.

Errors in the spectral index map can arise from zero-level or
temperature-scale errors at either frequency.  We deal with zero-level
errors first.  Taking 5 kK as a possible error in the 22 MHz data (see
the detailed discussion in Section 6.1) we estimate the effects on the
spectral index map. The large frequency separation between 22.25 and
408 MHz means that zero-level errors have relatively little effect:
the $\pm$5 kK error will change spectral index by $\pm$0.1 at the sky
minimum and by $\pm$0.01 on the brightest part of the Galactic
plane. We tested the effect of an error of $\pm$5 kK on the map of
Fig.~10 by computing maps with this error applied to the 22
MHz data in both senses. All the main features visible in Fig.~10 
remain in both new maps. When we discuss the spectral features which
we see in Fig.~10, we discuss only those which survive this test.

The effects of temperature scale errors are more difficult to
assess. Once again, the large frequency separation is an asset.  At
the zenith (declination 48.8\degr) we have an independent
determination of spectral index since we have not changed our data in
any way at that declination. We estimate that the probable error there
is 0.05, taking into account both systematic and random errors
(amounting to a 16\% difference in the temperature ratio
$T_{22}/T_{408}$), and we assign this error to the whole map. Further
errors at other declinations depend on the validity of the assumption
on which our calibration of the 22 MHz temperature scale is based, the
constancy of the spectral index of the Galactic emission over the
region 8$^h$ to 16$^h$, $-$28\degr\ to 80\degr. This can only be
tested with an extensive study of spectral index using data at a
number of frequencies, a study beyond the scope of this paper.  Note
that we have assumed the constancy of the {\it{differential}}
spectrum: we are assuming that the Galactic component of the emission
has constant spectral index over this region, which includes the
Galactic pole and lies mostly at latitudes higher than
$\sim$20\degr. The {\it{total}} spectrum includes an extragalactic
(presumably isotropic) component of emission, and the total spectrum
may still vary across this region, as it appears from Fig.~10 to do.

In the final analysis, the main value of our spectral index work is in
the assessment of {\it{differences}} in spectral index between regions
rather than in a precise determination of the spectral index of a
given region. In this spirit, we make the following observations.

\begin{list}{}{} \item[1.] In general, the map in Fig.~10 shows few
large scale variations.  The mean index away from the Galactic plane
is 2.47.  Within this area (latitude greater than $\sim$ 10\degr) the
mean index of extended regions varies from values of 2.41 in the area
of synchrotron minimum (10$^h$, 40\degr) to 2.54, near the position of
Loop III, both with an {\it rms} variation of $\sim$0.03.  On the
Galactic plane, the flatter spectrum region coincident with the
absorption trough at low longitudes is narrow in the latitude
dimension.  At all points on the Galactic plane it is notable that the
spectral index is relatively constant over most of the broader
synchrotron emission along the plane with no gradual decrease in
spectral index as latitude decreases.

\item[2.] The outline of the North Polar Spur at higher latitudes can
be seen in the spectral index map.  An integration over the area of
this arc shows its emission to have an index of 2.51, slightly steeper
than its surroundings by $\sim$0.03.
\end{list}

\section{Factors affecting the accuracy}

\subsection {Zero--level offsets}

The offsets of the T-T plots between 408 MHz and the final 22 MHz map
are in the range $\pm$10~kK at 22 MHz. The mean of 11 offset values at
intervals of 10\degr\ in declination from $+$79\degr~ to $-$21\degr~
is $-$5.1~kK with an {\it rms} deviation of 11.1~kK.  This mean offset
is of the same order as that derived for the T-T plot of zenith data
(see Fig.~1).  From these data alone, however, it is not
possible to decide the extent of the true offsets in either the 22~MHz
or the 408~MHz data.  The 408 MHz data is suspected of having offsets
as large as 2~K (Reich \& Reich \cite{reich88}) and may also have a
baselevel correction of 5~K. If the average spectral index between 22
and 408 MHz is 2.50, these values correspond to 2.8~kK and 7.2~kK at
22 MHz.

\subsection{Effects of antenna properties on the data}

In this section we examine the possible effects on our data of our
imperfect knowledge of the properties of the antennas. The antennas
used at 22 MHz and 408 MHz are of different types, and need to be
discussed separately.

If emission received in the sidelobes made a significant contribution
to antenna temperature, then errors would result. These would be
especially significant in the map of spectral index, because the
sidelobe responses of the two very different antennas would receive
emission from different parts of the sky. The greatest effect would
occur in measurements of those parts of the sky where the brightness
temperature is the lowest (around 9 hours, 30\degr) with emission from
the bright Galactic plane being received in the far sidelobes of the
antennas. As an example of such effects, Landecker \& Wielebinski 
(\cite{landecker}),
using the Parkes 64-m telescope at 150 MHz, found that about one third
of the antenna temperature at the sky minimum arose from sidelobe
contributions.

The response of the 22 MHz antenna is, in principle, completely
calculable from the geometry of the array, and from the phasing and
grading applied to the array elements. The angular
size of the main beam proved to be very close to the calculated value.
The net sidelobe solid angle should be zero, implying a beam
efficiency of 1.0. Sidelobe responses, apart from their effects near strong
sources, should not affect the measurement of the broad structure
which is the focus of this paper. Measurements of the antenna response
using the bright sources Cas A and Cyg A (Costain et al. \cite{costain}) bear out
this expectation. The dynamic range of these measurements is about 30
dB, determined by the ratio of the flux density of Cyg A (29,100 Jy)
to the confusion limit for the telescope (~30 Jy). Sidelobes above
this level were confined to the NS and EW planes (strictly a small
circle EW, depending on the phasing in declination) and were alternately 
positive and negative, as expected, and
close to the predicted amplitude.  Sidelobe response fell below the
detection limit within 10\degr\ of the main beam in the EW direction 
and below 1\% within 18\degr\ of the main beam in the NS
plane. These measurements verify our assertion that the performance of
the telescope at the zenith is well understood (Cas A and Cyg A pass
within 10\degr\ of the zenith at DRAO).

We are confident that the response of the antenna to the extended
background was as predicted near the zenith because of the linear
relationship between the 22 and 408 MHz brightness temperatures with
the expected  spectral index at that declination
(48.8\degr).  We know that the behaviour away from the zenith
departed from the expected response for the extended emission
features, but not for point sources. We tentatively attribute this 
to inadequately compensated mutual impedance effects between phased 
rows of dipoles in the array. At increasing zenith angles, these 
effects may have dominated the predicted response of individual dipoles 
above a reflecting screen.

The available measurements suggest, however, that the sidelobes of the
complete telescope (as opposed to the individual groups of radiating
elements) were still confined to the predicted regions, even at large
zenith angles. If we assume that the sidelobe level was all
positive and at the detection limit ($-$30 dB) in two 180\degr\ strips
in the EW and NS directions, each equal in width to the main beam,
then the beam efficiency would be 0.9, better than most reflector
antennas. However, this is very much a worst-case assumption, and it
is probably safe to conclude that the beam efficiency was $\geq$
0.95. Furthermore, the largest sidelobes lie in the NS plane, and,
when the main beam is measuring the coldest region of the sky, these
sidelobes do not intersect the bright Galactic plane. We therefore
feel justified in ignoring the effects of the sidelobe response of the
22 MHz telescope.

The 408 MHz data were not directly corrected for sidelobe contributions, but an
indirect correction was applied (Haslam et al. \cite{haslam82}). The 
absolutely calibrated survey at 404 MHz made by Pauliny-Toth \& Shakeshaft 
(\cite{pauliny-toth}) with a beam of
about 7.5\degr~ was used to establish both the zero-level and the temperature
scale of the 408 MHz survey data. Since the 404 MHz survey {\it{was}} corrected
for sidelobe contributions, using it as a reference for the later survey 
roughly corrected those measurements for sidelobe contributions. The technique
is valid in this case since the relationship between the measured antenna 
temperature and the sidelobe correction is, to a good approximation, linear.
Checks of the effectiveness of this procedure were made subsequently 
by Lawson et al. (\cite{lawson}) and by Reich \& Reich (\cite{reich88}) 
who convolved the 408 MHz data to the broad beams of horns and other 
low--sidelobe antennas used by Webster (\cite{webster}) and Sironi 
(\cite{sironi}) to measure the Galactic emission at 408 MHz.  In all cases 
the comparisons were satisfactory, indicating that sidelobe effects had 
been effectively removed from the 408 MHz data.


\section{Discussion}

\subsection{Components of extended Galactic emission}

Beuermann et al. (\cite{beuermann}) have used the 408 MHz all-sky survey
(Haslam et al. \cite{haslam82}) to produce a three-dimensional model
of the Galactic radio emission using an unfolding procedure.  In this
model the Galaxy consists of a thick non-thermal radio disk in which a
thin disk is embedded.  The thick disk exhibits spiral structure, has
an equivalent width of $\sim$3.6~kpc at the solar radius and accounts
for $\sim$90\% of the diffuse 408 MHz emission.  Emission extends to
at least 15~kpc from the Galactic centre, at which radius the thick
disk has an equivalent width near 6~kpc.  The {\it thin} disk, by
comparison, appears in the model as a mixture of thermal and
non-thermal emission also with spiral structure, but with an
equivalent width of $\sim$370~pc, similar to that of the \ion{H}{i}
disk and of the distribution of \ion{H}{ii} regions in the inner
Galaxy.

Our comparison of the 22~MHz and 408~MHz maps shows a remarkable
constancy of spectral index in the extended emission corresponding to
the thick disk component over our full range of longitudes from
$\sim$0\degr\ to $\sim$240\degr.  The principal departures from this
general tendency are {\it{(i)}} the slightly flatter spectral index in
a broad area in the region of minimum Galactic emission at high
latitudes toward the longitude of the Galactic anticentre and
{\it{(ii)}} the somewhat steeper indices near Loop III and the outer
rim of the NPS.  In a similar comparison of the 408~MHz map with a map
of 1420~MHz emission, Reich \& Reich (\cite{reich88}) also note these
general features.  However, we see no indication in the lower
frequency range for the steeper spectra seen by Reich \& Reich in
regions on the plane both near the Galactic centre and near longitude
130\degr.  This suggests that any steepening of the spectra in these
regions must be a higher frequency phenomenon with spectral curvature
above 408~MHz.
  
Details of the spectral index variations associated with the loops of
emission also differ in the two frequency ranges.  Fig.~10
shows a slightly steeper index (by $\sim$0.03) for a substantial part
of the arc forming the outer edge of the NPS.  This contrasts with the
408--1420MHz comparison (Reich \& Reich \cite{reich88}) which shows a
steeper index in a relatively broad arc on the part of the NPS closest
to the Galactic plane. Neither study indicates a difference between
the spectral index of emission within the loop of the NPS and that
outside the loop.  The NPS has variously been considered as a nearby,
very old supernova remnant (e.g. Salter \cite{salter}) and as a local
magnetic ``bubble'' (Heiles \cite{heiles}).

\subsection{Absorption at lower longitudes}

We have noted that at longitudes less than $\sim$40\degr~ there exists
a continuous trough of absorption along the Galactic plane.  We
illustrate this in Fig.~11 which shows a map of the ``quasi
optical depth'' at 22 MHz calculated from a comparison with the 408
MHz map on the assumption that the absorption is due entirely to cool
ionized gas on the near side of the emission.  (We define quasi
optical depth, ${\tau}'$, by the relation ${\tau}' = ln\left(T_{408}
(408/22)^{{\beta}'}/{T_{22}}\right)$, where ${\beta}'$ is the mean
spectral index of the emission off the Galactic plane).  This
represents an underestimate of the true optical depth of the absorbing
gas since ({\it i}) a proportion of the non-thermal emission will be
on the near side of some absorption and, ({\it ii}) the kinetic
temperature of the thermal gas will lessen the apparent depth of the
absorption.  A more accurate estimate of the true optical depth would
require a modelling of the intermixed emission and absorption
components which is beyond the scope of this paper.  Nonetheless, it
is obvious from Fig.~11 that the full angular width of the absorbing
region is less than 3\degr~ which, at an assumed mean distance of
4~kpc, corresponds to a thickness of less than 250~pc.  Thus, it is
apparent that the absorption corresponds to the ``thin disk
component'' of emission identified by Beuermann et al
(\cite{beuermann}) as comprising the known disks of \ion{H}{ii}
regions, diffuse thermal continuum emission, diffuse recombination
line emission and the distribution of atomic hydrogen.

The extended absorption in the plane in the region of Cygnus between
longitudes 70\degr\ and 90\degr\ is also shown in Fig.~11.
Note that the region appears at least twice as extensive in latitude
as the continuous trough, probably because much of the absorbing gas
is at distances of 1~kpc or less.

\subsection{Non-thermal emissivities in the plane}

Several of the discrete \ion{H}{ii} regions which appear in absorption at 22 MHz and
which are listed in Table 2 can be used to estimate the emissivity of local synchrotron
emission. We have calculated the emissivities for eight \ion{H}{ii} regions at 
well--determined distances, which are sufficiently extended compared to the 
observing beam to ensure that only thermal radiation from the ionized gas 
and foreground non-thermal radiation contribute to the measured emission.  
An assumed contribution from the opaque ionized gas of 6000K was
subtracted from the brightness temperature in the depression and the
result divided by the distance to the \ion{H}{ii} region. The values
of emissivity are presented in Table 3.

\begin{table*}
\caption[]{Synchrotron emissivities in the directions of \ion{H}{ii} regions}
\label{tabl-3}
\begin{center}
\begin{tabular}{lllll}
\hline
Galactic Coordinates & Region & Distance & Foreground emission & Emissivity \\
~ & ~ & {\it parsecs} & {\it kK} & {\it K/pc} \\
\hline
$6.3+23.5$ & Sh~27 & 170$^{\mathrm{a}}$ & 27.0 & 159 \\
$85.5-1.0$ & Sh~117 & 800$^{\mathrm{b}}$ & 17.4 & 21.8 \\
$99.3+3.7$ & Sh~131 & 860$^{\mathrm{c}}$ & 37.4 & 43.5 \\
$118.5+6.0$ & NGC7822 & 840$^{\mathrm{c}}$ & 30.1 & 35.8 \\
$134.8+0.9$ & IC1805 & 2200$^{\mathrm{c}}$ & 46.0 & 20.9 \\
$160.1-12.3$ & Sh~220 & 400$^{\mathrm{c}}$ & 23.9 & 59.8 \\
$195.1-12.0$ & Sh~264 & 400$^{\mathrm{c}}$ & 20.4 & 51.0 \\
$202.9+2.2$ & Sh~273 & 800$^{\mathrm{d}}$ & 23.1 & 28.9 \\
\hline
\end{tabular}
\end{center}
\begin{list}{}{}
\item[$^{\mathrm{a}}$] Georgelin et al. (\cite{georgelin73})
\item[$^{\mathrm{b}}$] Georgelin (\cite{georgelin75})
\item[$^{\mathrm{c}}$] Humphreys (\cite{humphreys})
\item[$^{\mathrm{d}}$] Turner (\cite{turner})
\end{list}

Note: The electron temperature in the \ion{H}{ii} regions is assumed
to be 6000 K.

\end{table*}

In the longitude range 85\degr\ to 205\degr, six \ion{H}{ii} regions
are at distances from 400--900 pc and the values of 22 MHz
emissivity\footnote{A volume emissivity per unit line-of-sight of
1~Kpc$^{-1}$ is equal to
4.93$\times$10$^{-42}$Wm$^{-3}$Hz$^{-1}$sr$^{-1}$ at 22 MHz} range
from 21~Kpc$^{-1}$ to 60~Kpc$^{-1}$ with a mean of 40.1~Kpc$^{-1}$.
Excluding two \ion{H}{ii} regions, Sh220 and Sh264, which are more
than 10\degr~ off the plane, but including IC1805, at a distance of
2.2 kpc, we find a mean emissivity of 30.2~Kpc $^{-1}$ with an {\it
rms} of 9.6 ~Kpc$^{-1}$.  These emissivities are comparable with
similarly derived emissivities tabulated (at 10MHz) by Rockstroh \&
Webber (\cite{rockstroh}).  In addition, our value in the direction of
IC1805, 20.9~Kpc$^{-1}$, is close to the value of 18~Kpc$^{-1}$
obtained by Roger (\cite{roger69a}) using a detailed modelling of 22
and 38 MHz data for the IC1805--IC1848 complex.

However, there is a problem reconciling a mean value of local emissivity of 
30~Kpc$^{-1}$ with the model of Galactic emission of Beuermann et al. 
(\cite{beuermann}) which assumes a lesser value of 15~Kpc$^{-1}$ 
(11~Kkpc$^{-1}$ at 408 MHz) at the solar radius.  If we take the value
of the brightness temperature at the Galactic poles (27~kK), subtract an 
extragalactic component of 6~kK (Lawson et al. \cite{lawson}) and divide 
by the model's 
half--equivalent--width of 1.8~kpc, we derive a mid-plane emissivity of
only 11.7~Kpc$^{-1}$, almost a factor of 3 less than our measured mean
value.  To reconcile our measurement with the model, one or more of the
following must apply: {\it (i)} our measured local mean emissivity is greater
than the typical value at the solar radius; {\it (ii)} the equivalent
width of the ``thick--disk'' component is locally less than the model
predicts; {\it (iii)} the extragalactic component of the polar emission 
is less than is estimated from extrapolations of extragalactic source counts 
at higher frequencies; and/or {\it (iv)} a zero--level correction should 
be added to the 22 MHz brightness temperatures. With regard to the 
extragalactic component of emission, we note that estimates are usually
derived from source count (``log N--log S'') analyses at frequencies 
above 150~MHz (e.g. Lawson\cite{lawson}), extrapolated with an assumed
spectral index $\beta \approx$2.75.  Analyses of source counts at substantially
lower frequencies are needed for accurate estimates of the extragalactic component. 
We noted in Section 6.1 the possibility of a zero--level correction as
indicated by T-T plot comparisons with 408~MHz data.  In this regard, it 
is interesting to note that very low resolution measurements with scaled 
antennas at several low frequencies (Bridle \cite{bridle})
predicted a brightness temperature at 22 MHz in the area of the North
Galactic Pole 4~kK higher than our value.   This is of the same magnitude
and sense as the offset suggested by the T-T plot analysis.

We note the unusually high emissivity derived for the direction toward
$\zeta$-Oph (Sh27), a relatively nearby complex some 23\degr~ above
the plane at the longitude of $\sim$6\degr.  Emission from this
direction may include components from the North Polar Spur and from a
minor spur that is most prominent near $l=6$\degr, $b=14$\degr, both
of which may be foreground features. Also, it is possible that this
somewhat diffuse region is not completely opaque at 22 MHz, in which
case an unknown amount of background emission may contribute a
spurious component to the emissivity.  
\bigskip

\noindent{\it Acknowledgements.} We are indebted to several colleagues
for their assistance in collecting and processing the observational
data, and we particularly thank J. D. Lacey, J.H. Dawson and
D.I. Stewart.  We are also grateful to Dr. J.A. Galt for his
encouragement at various stages of this project.


{\noindent{\bf{FIGURE CAPTIONS}}}

{\noindent{\bf{Fig. 1}}} A T-T plot of 22 MHz and 408 MHz data for
declination 48.8\degr and all right ascensions except (a) for the
range 20$^h$ to 22$^h$ and (b) small regions around bright point
sources.

{\noindent{\bf{Fig. 2}}} A map of the emission at 22~MHz, continued in
Figs. 3 to 6.  Contours of brightness temperature are at the following
levels in kilokelvins: 14, {\bf19}, 24, 29, {\bf34}, 41, 48, {\bf55},
65, 75, {\bf85}, 100, 115, {\bf130}, 150, 170, {\bf190}, 220, 250.
Levels in bold have thick contour lines and are labelled.  Regions
affected by sidelobes of four strong sources are blanked out. The
superimposed grid shows Galactic co-ordinates in steps of 30\degr\ in
$l$ and $b$, labelled only where grid lines intersect the right-hand
side and the top of the figure.

{\noindent{\bf{Fig. 3}}} Fig. 2 continued with the same contours and
grayscale, showing the Galactic anticentre. A region around Tau A has
been blanked out.

{\noindent{\bf{Fig. 4}}} Fig. 2 continued with the same contours and
grayscale. A region around Cas A has been blanked out.

{\noindent{\bf{Fig. 5}}} Fig. 2 continued with the same contours and
grayscale, showing the central regions of the Galaxy. Note the deep
absorption trough along the plane near the Galactic centre. The North
Polar Spur rises from the Galactic plane at $l$ = 30\degr. A region
around Cyg A has been blanked out.

{\noindent{\bf{Fig. 6}}} Fig. 2 continued with the same contours and
grayscale. The large circular feature is the North Polar Spur. A
region around Vir A has been blanked out.

{\noindent{\bf{Fig. 7}}} The 22~MHz emission in Galactic coordinates in
two segments with positions of prominent Galactic sources indicated.
Contours, at the same levels shown in Figs. 2 to 6, are indicated with
a bar scale.  The superimposed grid shows equatorial co-ordinates
(J2000) in steps of 2$^h$ in right ascension and 30\degr\ in
declination.

{\noindent{\bf{Fig. 8}}} Fig. 7 continued.

{\noindent{\bf{Fig. 9}}} An Aitoff projection of the 22~MHz emission in
Galactic coordinates.  The Galactic centre is at the map centre and
grids are at 30\degr\ intervals in longitude and latitude, positive to
the left and upwards respectively.  Note the arc of the North Polar
Spur rising from the plane near longitude $+$30\degr.

{\noindent{\bf{Fig. 10}}} A map of the spectral index from 22~MHz to
408~MHz in shaded grey levels as indicated by the bar scale.  Regions
near four strong sources are blanked out. The superimposed grid shows
Galactic co-ordinates in steps of 30\degr\ in $l$ and $b$.

{\noindent{\bf{Fig. 11}}} The ``quasi optical depth'' at 22~MHz along
the Galactic plane in the first quadrant, from a comparison of the
408~MHz and 22~MHz emissions, assuming all absorbing (thermal) gas is
on the near side of the background synchrotron emission.  Contours are
at optical depths of 0.4, 0.8, 1.2, 1.6 and 2.0.


\begin{thebibliography}{}

  \bibitem[1997]{alvarez} Alvarez, H., Aparici, J., May, J., Olmos, F.,
   1997, A\&AS, 124, 315

  \bibitem[1985]{beuermann} Beuermann, K., Kanbach, G., Berkhuijsen, E.M.,
   1985, A\&A, 153, 17

  \bibitem[1967]{bridle} Bridle, A.H., 1967, MNRAS, 136, 219

  \bibitem[1976]{caswell} Caswell, J.L., 1976, MNRAS, 177, 601

  \bibitem[1969]{costain} Costain, C.H., Lacey, J.D., Roger, R.S., 1969, 
   IEEE Trans-AP 17, 162

  \bibitem[1982]{ellis} Ellis, G.R.A., 1982, Aust. J. Phys. 35, 91

  \bibitem[1975]{georgelin75} Georgelin, Y.M., 1975, th\`ese de doctorat,
   Universit\'e de Provence

  \bibitem[1973]{georgelin73} Georgelin, Y.M., Georgelin, Y.P., Roux, S.,
   1973, A\&A, 25, 337

  \bibitem[1969]{hamilton} Hamilton, P.A., Haynes, R.F., 1969, Aust. J. Phys.
   22, 839

  \bibitem[1982]{haslam82} Haslam, C.G.T., Salter, C.J., Stoffel, H., Wilson, 
   W.E., 1982, AAS, 47, 1

  \bibitem[1998]{heiles} Heiles, C. 1998, The Magnetic Field near the Local
   Bubble. In: Breitschwerdt, D., Freyberg, M.J., Tr\"{u}mper, J. (eds.)
   Proc. IAU Coll. 166, The Local Bubble and Beyond. Springer--Verlag
   (Lecture Notes in Physics, Vol 506)

  \bibitem[1978]{humphreys} Humphreys, R.M., 1978, ApJS, 38, 309

  \bibitem[1987]{landecker87} Landecker, T.L., Vaneldik, J.F., Dewdney, P.E., 
   Routledge, D., 1987, AJ, 94, 111

  \bibitem[1970]{landecker} Landecker, T.L., Wielebinski, R., 1970, 
   Aust. J. Phys. Suppl. 16, 1

  \bibitem[1987]{lawson} Lawson, K.D., Mayer, C.J., Osborne, J.L., 
   Parkinson, M.L., 1987, MNRAS, 225, 307

  \bibitem[1973]{milogradov} Milogradov-Turin, J., Smith, F.G., 1973, MNRAS 161, 269

  \bibitem[1962]{pauliny-toth} Pauliny-Toth, I.I.K., Shakeshaft, J.R., 
   1962, MNRAS, 124, 61

  \bibitem[1997]{pineault} Pineault, S., Landecker, T.L., Swerdlyk, C.M., Reich, W.,
   1997, AA, 324, 1152

  \bibitem[1988]{reich88} Reich P., Reich W., 1988, AAS, 74, 7

  \bibitem[1978]{rockstroh} Rockstroh, J.M., Webber, W.R., 1978, ApJ, 224, 677

  \bibitem[1969]{roger69a} Roger, R.S., 1969, ApJ, 155, 831

  \bibitem[1973]{roger73} Roger, R.S., Bridle, A.H., Costain, C.H., 
   1973, AJ, 78, 1030

  \bibitem[1976]{roger76} Roger, R.S., Costain, C.H., 1976, AA, 51, 151

  \bibitem[1969]{roger69b} Roger, R.S., Costain, C.H., Lacey, J.D., 
   1969, AJ, 74, 366

  \bibitem[1986]{roger86}Roger, R.S., Costain, C.H., Stewart, D.I., 
   1986, AAS, 65, 485

  \bibitem[1983]{salter} Salter, C.J., 1983, Bull. Astr. Soc. India 11, 1

  \bibitem[1974]{sironi} Sironi, G., 1974, MNRAS, 166, 345

  \bibitem[1976]{turner} Turner, D.G. 1976, ApJ, 210, 65

  \bibitem[1962]{turtle} Turtle, A.J., Baldwin, J.E., 1962, MNRAS 124, 459

  \bibitem[1975]{webster} Webster, A.S., 1975, MNRAS, 166, 355

\end{thebibliography}
\end{document}